\documentclass[aps,twocolumn,showpacs,amsmath,amssymb]{revtex4}
\usepackage{graphicx}% Include figure files
\usepackage{dcolumn}% Align table columns on decimal point
\usepackage{bm}% bold math
\usepackage{hyperref}

\begin{document}

\title{Traveling Bubbles and Vortex Pairs within Symmetric 2D Quantum Droplets}
\author{Angel Paredes$^1$, Jose Guerra-Carmenate$^{2,1}$, Jose R. Salgueiro$^1$, Daniele Tommasini$^1$ and Humberto Michinel$^1$}
\affiliation{$^1$Instituto de F\'\i sica e Ciencias Aeroespaciais (IFCAE), 
Universidade de Vigo. Campus de As Lagoas, E-32004 Ourense, Spain.\\
$^2$Instituto Universitario de Matem\'atica Pura y Aplicada (IUMPA),
Universitat Polit\`ecnica de Val\`encia. Edificio 8E, 46022 Valencia, Spain.\\
}

\begin{abstract}
%----------------------------   ABSTRACT  ------------------------------------
We disclose a class of stable nonlinear traveling waves moving at specific constant velocities within symmetric two-dimensional quantum droplets. We present a 
comprehensive analysis of these traveling bubbles and identify three qualitatively distinct regions within the one-parameter family of solutions, classified by velocity: 
(i) well-separated phase singularities at low velocity, (ii) singularities within the same density dip at intermediate velocity,
 and (iii) rarefaction pulses without singularities at higher (subsonic) velocities.
Then, we generalize the discussion to unstable cases, incorporating higher order vortex-antivortex pairs and
arrays of vortices that move cohesively with a common velocity within the fluid.
In all cases, we provide analytic approximations that aid the understanding of the results in different regimes.

\end{abstract}
%-----------------------------------------------------------------------------
\maketitle
%--------------------------------------------------------------------------------

\section{Introduction}

Dark solitons \cite{kivshar1998dark} are 
localized wave perturbations that preserve their form as they 
propagate through a nonlinear medium \cite{hasegawa1973b}. 
In contrast to their bright version \cite{shabat1972exact,hasegawa1973a}, which 
manifests as amplitude peaks, dark solitons are characterized by dips or troughs 
in a coherent field \cite{Swartzlander1991,Saqlain2023}. Within the framework of nonlinear 
Schr\"odinger equations (NLSE) \cite{rabinowitz1992class,blow1985multiple,kevrekidis2015}, these 
solutions emerge when self--repulsive interactions are present and have been 
observed in various physical contexts \cite{kivshar1989}, such as optics \cite{weiner1988experimental}, 
Bose--Einstein condensates 
(BECs) \cite{becker2008oscillations,weller2008experimental,kengne2021spatiotemporal}, 
and superfluids \cite{roberts2001nonlinear,Buelna2019}, among 
others \cite{almand2014,uchida2010,pismen1999vortices,paredes2019vortex,nugaev2020review}.

Vortex solitons \cite{donadello2014observation,swartzlander1992,malomed2019vortex} can be considered as a 
particular case of dark solitons \cite{desyatnikov2005optical} taking the form of 
steady dark spots (in 1+2 dimensions) or lines and rings (in 3D configurations)
corresponding to phase singularities around which there is a rotating  flow \cite{coullet1989,quiroga1997}. 
The relationship between these types of nonlinear waves has been analyzed extensively 
in the literature \cite{kivshar2000,malomed2019vortex,shen2019,paredes2022vortex} and it 
is well known that, under certain conditions, a dark soliton can transition into a 
vortex \cite{tikhonenko1996,anderson2001} due to nonlinear perturbations such as the 
so--called {\it snake instability} \cite{kuznetsov1986soliton}. It has also been noticed that 
vortices interact through their ``topological charge" \cite{paz2005,paredes2023polygons}, which is the integer 
number of phase windings around the singularity. Thus, vortex and dark solitons  move 
through nonlinear media affected by topological interactions and amplitude 
gradients \cite{rozas1997,dreischuh2006,smirnov2015scattering,groszek2018motion,de2023method}. 
Being low density regions within a fluid, these 
kinds of solutions represent different examples of ``bubbles'' 
\cite{zeng2021flat,barashenkov1988soliton,berloff2004vortex,katsimiga2023}, 
whose analysis will be the central goal of this work.

Specifically, we will deal with bubbles within quantum droplets (QDs), a particular case 
of BECs stabilized by a quantum effect first discussed
by  Lee, Huang and Yang (LHY) \cite{lee1957}. Usually, the LHY effects are negligible, but Petrov showed that in appropriately
tailored binary mixtures, they can stabilize the condensate against collapse,
yielding a stable self-trapped atom cloud, namely a QD \cite{petrov2015quantum}.
This prediction was subsequently demonstrated in the laboratory  \cite{Schmitt2016,cabrera2018,semeghini2018}, sparking an
intense research effort both from the theoretical and experimental perspectives, see \cite{Bottcher2021,luo2021new,guo2021new,malomed2021} for reviews.

QDs support different kinds of form-preserving dark excitations.
Vortex solutions within 2D and 3D QDs have been extensively studied, see, e.g., \cite{li2018two,kartashov2018three,paredes2024variational}, including
cases with different vorticities for the different components of the mixture \cite{huang2022}.
   Codimension one dark stripes within a 2D QD were analyzed in \cite{bougas2024stability}.
  They were proved to be subject to snake instability, with similar behavior to cases in different physical situations \cite{Kuznetsov1988,swartzlander1992,Kuznetsov2022}.
  There have been several interesting works concerning dark excitations in 1D QDs:
  unstable bubbles have been
  discussed in \cite{katsimiga2023}, the effect of a harmonic trap was analyzed in \cite{katsimiga2023b}, beyond mean-field effects and a crossover
   from a dark soliton to the dark quantum droplet were studied in \cite{edmonds2023} and dispersive shock waves in \cite{Chandramouli2024}.
  
The aim of this work is to explore a novel class of dark traveling waves within quantum droplets (QDs), building upon and extending the findings of \cite{jones1982motions} to this context. Unlike individual vortices, the bubbles we analyze can exist within a uniform-density background while preserving boundary conditions at infinity. This makes them a compelling example of localized structures in matter waves \cite{mihalache2021localized} and offers new insights into the possible excitations of flat-top QDs.
Specifically, we focus on the symmetric two-dimensional case and employ the mathematical model developed by Petrov and Astrakharchik \cite{petrov2016ultradilute}, which we summarize in Section \ref{sec:math}. This model incorporates nonlinear potential terms with logarithmic corrections, leading to a density-dependent interaction: attractive interspecies interactions prevail at low densities, while repulsive intraspecies interactions dominate at higher densities.
Through a combination of analytical and numerical methods, we obtain the main results of this study, which can be summarized as follows::
 
\begin{itemize}

\item In Section \ref{sec:stable}, we compute and analyze a one-parameter family of stable, shape-preserving traveling wave solutions. The parameter governing this family is the velocity, 
$0<U<U_0$, where $U_0$ denotes the speed of sound. We derive three virial identities that these solutions satisfy and provide analytical approximations in the limits 
$U \to 0$ and $U \to U_0$. Additionally, we compute numerical approximations for the entire family, revealing three qualitatively distinct regimes: (i) well-separated vortex-antivortex 
pairs at low velocities, (ii) vortex-antivortex pairs within an elongated density trough at intermediate velocities, and (iii) rarefaction pulses without phase singularities at higher velocities.  

\item We then explore two other one-parameter families of solutions: vortices with higher topological charge (Section \ref{sec:largel}) and multi-vortex structures that propagate cohesively 
while preserving the network configuration (Section \ref{sec:linea}). Although the solutions discussed in Sections \ref{sec:largel} and \ref{sec:linea} are ultimately unstable, we demonstrate 
that they can persist for relatively long times, making them intriguing excitations of QDs.  

\end{itemize}
 
In section \ref{sec:conclusion}, we discuss the results and comment on possible lines of future research.

\section{Mathematical model}
\label{sec:math}

 In dilute BECs of alkali atoms, the dynamics of the coherent ultracold
gas is described by a Gross-Pitaevskii equation (GPE) that incorporates atom-atom interactions \cite{dalfovo1999theory}.
This model can be further enriched by the LHY corrections \cite{lee1957}, that modify the interactions and 
the ground state energy. We deal here with a binary mixture
with a strong external trapping along one spatial dimension ($z$), so the condensate takes the shape of
a pancake and the system becomes effectively two-dimensional.
Petrov and Astrakharchik provided an accurate mean-field description 
of such a binary system \cite{petrov2016ultradilute}, in which an effective nonlinear potential with a logarithm accounts for the 
attractive interspecies  and the repulsive intraspecies interactions. In the simplest and most studied case of equal densities of 
the two atomic species and symmetric self--interactions given by scattering lengths 
$a_{\uparrow\uparrow} = a_{\downarrow\downarrow} = a$, the system of two coupled GPEs can 
be reduced to a single one describing simultaneously the dynamics of the two species with a
single wavefunction, reading  \cite{petrov2016ultradilute}:
\begin{equation}
i\hbar \partial_{\tilde t} \tilde \Psi \! = \!- \! \frac{\hbar^2}{2 m} \tilde \nabla^2 \tilde \Psi 
\! + \!\frac{8 \pi \hbar^2}{m  \ln^2 (a_{\uparrow\downarrow}/a)}
\ln\! \left( \frac{|\tilde \Psi|^2}{n_0 \sqrt{e} } \!\right)\! |\tilde \Psi|^2 \tilde \Psi,
\label{eqdim}
\end{equation}
where we use tilded symbols for dimensionful quantities. $m$ is the boson mass,
$\tilde \nabla^2=\partial_{\tilde x}^2+ \partial_{\tilde y}^2$ is the two-dimensional Laplacian, $a_{\uparrow\downarrow}$, $a$ are respectively the 
inter--species and intraspecies scattering lengths and $n_0$ is the 
equilibrium density. 
 It is convenient to introduce a rescaling 
 to adimensional variables for the wavefunction ($\Psi$), time ($t$)
and space ($x,y$). 
\begin{eqnarray}
\tilde \Psi &=& e^\frac14 n_0^\frac12 \Psi ,\nonumber \\
\tilde t &=& \frac{m\,\ln^2 (a_{\uparrow\downarrow}/a)}{8\pi \hbar n_0 \sqrt{e}} t,  \nonumber\\
(\tilde x,\tilde y) &=& \frac{\ln (a_{\uparrow\downarrow}/a)}{4 e^\frac14 \sqrt{n_0 \pi}} (x,y),
\end{eqnarray}
which leads to: 
\begin{equation}
i\partial_{ t}  \Psi = -  \nabla^2  \Psi + \left[ |\Psi|^2\ln  (|\Psi|^2)\right]  \Psi .
\label{eqadim}
\end{equation}
Eq. (\ref{eqadim}) is the starting point of our analysis.
The form of the nonlinear potential is analogous to the well--known 
Shannon's expression for information entropy \cite{shannon1948} and therefore we will refer 
to this dependency as ``Shannon--type" nonlinearity. 
It is worth mentioning that the form of the potential in Eq. (\ref{eqadim}) changes in the case of thick pancakes with
looser trapping along $z$ \cite{lin2021two}. We will not study that case in the present contribution although we
would expect qualitatively similar results to those presented in the following sections.

Eq. (\ref{eqadim}) supports self-trapped stable solutions \cite{petrov2016ultradilute}.
 For large atomic number, such matter waves tend to have a nearly constant amplitude
$|\Psi| \approx \psi_{cr}$ in their region of support \cite{astrakharchik2018dynamics,pathak2022dynamics}, and are usually called
{\it flat-top}.
We refer to $\psi_{cr}$ as the critical amplitude and to $\psi_{cr}^2$ as the critical density.
 These states behave like an incompressible liquid-like fluid with approximately constant density and surface tension \cite{paredes2024variational}. 
On the other hand, the simplest solution of Eq. (\ref{eqadim}) is 
$\Psi = e^{-i \mu t} \psi_{\mu}$
with real constants $\mu$ and $\psi_\mu$ that satisfy 
$\mu = \psi_\mu^2 \ln (\psi_\mu^2)$. The plane wave with $\psi_\mu=\psi_{cr}$ can be called the critical plane wave.
It is of particular interest because of its connection to finite sized droplets, being their asymptotic limit as the number of atoms grows.

We are interested in dark excitations within this critical plane wave background and, therefore, 
we will work in the following with boundary condition:
\begin{equation}
\lim_{r\to\infty} \Psi = \psi_{cr} e^{- i\mu_{cr}t}\,,\ \ {\rm with}\ \ \psi_{cr} = e^{-\frac14}, \ \  \mu_{cr} = -\frac{1}{2\sqrt{e}},
\label{boundary_cond}
\end{equation}
where we have inserted the appropriate values of $\psi_{cr}$, $\mu_{cr}$ for the convention of Eq. (\ref{eqadim}),
see \cite{paredes2024variational}. The symbols
$r,\theta$ stand for the polar coordinates in the $x,y$ plane.

\begin{figure}[h!]
\begin{center}
\includegraphics[width=\columnwidth]{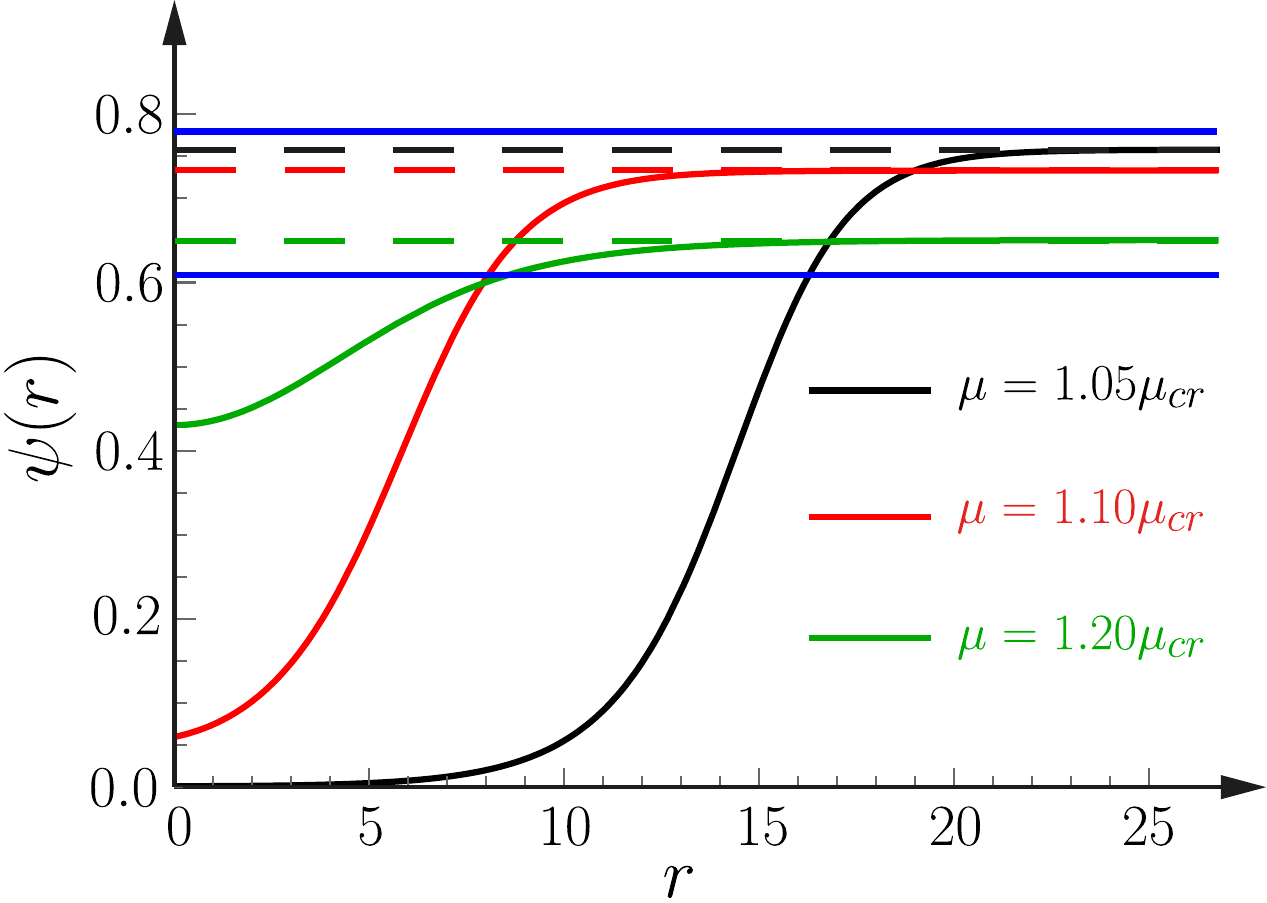}
\end{center}
\caption{Some sample profiles $\psi(r)$ for  quiescent unstable bubble solutions  in the range $\mu_{cr}  > \mu > - e^{-1}  $. 
 Horizontal dashed lines correspond to the asymptotic value $\psi_\mu = \lim_{r\to \infty} \psi(r)$, given in each case by the
equation $\mu= \psi_\mu^2 \ln (\psi_\mu^2)$. The horizontal solid lines represent the limiting
values of $\psi_\mu$ for the existence of solutions: $\psi_{cr}=e^{-\frac14} \approx 0.779$ and
$e^{-\frac12} \approx 0.607$, corresponding respectively to $\mu=\mu_{cr}$ and $\mu=- e^{-1}$.}
\label{fig1}
\end{figure}

Before discussing the traveling solutions in the sections \ref{sec:stable}, \ref{sec:largel}, \ref{sec:linea}, it is natural to wonder whether there can exist quiescent bubbles.
Taking the simple ansatz $\Psi = \psi(r) e^{-i\mu t}$, we have checked that there are bubble solutions
for the range $\mu_{cr}  > \mu > - e^{-1}  $. In particular, there are no solutions for the critical
background $\mu=\mu_{cr}$.
 The minimum value of the atom density
 $\psi(r=0)^2$ is always non-zero, although it approaches zero in the $\mu \to \mu_{cr}$ limit.
 Some examples are shown in Fig. \ref{fig1}. All of them are unstable. 
  The instability arises because the dip is quickly filled by the surrounding atoms, restoring the uniform background while generating ripples that propagate outward during this dynamic process. Note that, unlike in the case of vortices, there is no angular momentum preventing the occupation of the low-density region.
  These states can be considered as the two-dimensional
 counterpart of the one-dimensional bubbles described in \cite{katsimiga2023}. 
 They are the generalization to the QD framework of the bubbles of  \cite{barashenkov1988soliton}.

%%%%%%%%%%%%%%%%%%%%%% FIGURE 1 %%%%%%%%%%%%%%
%%%%%%%%%%%%%%%%%%%%%%%%%%%%%%%%%%%%%%%%%%

We close this section with two comments that will be useful for future reference.
First, we note that the long wavelength speed of sound in this fluid is:
\begin{equation}
U_0 = e^{-\frac14}\,.
\label{sound_speed}
\end{equation}
This value can be readily found by a standard procedure, perturbing the critical plane wave $\Psi=\left(\psi_{cr} + \rho(t, \mathbf{x}) + i \sigma(t, \mathbf{x}) \right) e^{- i \mu_{cr} t}$,
and solving for the linearized perturbations discarding fourth-order spatial derivatives.

We will also use the fact that there are vortex solutions \cite{li2018two,paredes2024variational}
of the form:
\begin{equation}
\Psi_{vort}= e^{-i\mu_{cr} t} \psi_l (r) e^{\pm i \,l\, \theta}\,,
\label{psi_vort}
\end{equation}
where  $\pm l$ is the topological charge.  The
profile of the vortex cores $\psi_l (r)$ depends on $|l|$ and can be found numerically imposing
$\lim_{r\to \infty} \psi_l(r)=\psi_{cr}$. Notice that (\ref{psi_vort}), having net vorticity,  cannot satisfy the boundary condition
(\ref{boundary_cond}). However, it will be useful as a building block for the discussion in the following sections.

\section{Stable traveling bubbles}
\label{sec:stable}

After showing that there are no static bubbles in the critical background, we now turn to configurations
moving with constant speed $U$, generalizing results of \cite{jones1982motions,bethuel2009} to the QD case.
 In
order to be form-preserving, $\Psi$ must take the form
$\Psi (t,x,y) = e^{-i\mu_{cr}t} \psi (x-U t, y )$, where we have chosen, without loss of generality, that
the motion is along the $x$-direction. 
Inserting this ansatz in Eq. (\ref{eqadim}), we find the following differential equation: 
\begin{equation}
i\,U\,\partial_x \psi = \nabla^2 \psi - \left(|\psi|^2 \ln (|\psi|^2) + \frac{1}{2\sqrt{e}}\right) \psi\,.
\label{eq_caviton}
\end{equation}
This expression holds for \( t = 0 \) and can be generalized to any \( t \) by substituting \( x \) with the comoving coordinate \( x - Ut \). For notational simplicity, we continue to express everything in terms of \( x \), although it should be understood as the comoving coordinate in the more general context.
The boundary condition (\ref{boundary_cond})  is:
\begin{equation}
\lim_{r\to \infty} \psi(x,y) = \psi_{cr}\,.
\label{boundary_cond_2}
\end{equation}
We can define energy and momentum as:
\begin{eqnarray}
E&= & \int \left[ |\nabla \psi|^2 + \frac14 |\psi|^4 \Big(2\ln (|\psi|^2) -1)\Big) + \frac{1}{2\sqrt{e}}|\psi|^2 \right] dx dy\nonumber\\
p&=&\frac{1}{2i} \int  \left[ (\psi^* - \psi_{cr}) \partial_x \psi - (\psi -\psi_{cr} ) \partial_x \psi^*  \right] dx dy \,.
\label{pE_defs}
\end{eqnarray}
The plane wave is simply $\psi=\psi_{cr}$ and has vanishing $E$ and $p$.
Suppose that we have some $\psi(x,y)$ that solves the system (\ref{eq_caviton}), (\ref{boundary_cond_2}).
If we take a small perturbation $\psi \to \psi + \delta \psi$, it is straightforward to check that 
$\delta p = -i \int \left[\delta \psi^* \partial_x \psi - \delta \psi \partial_x \psi^*\right] dxdy$ and
$\delta E = U \delta p$. Therefore, along any family of solutions, we have the usual expression for the group velocity in terms
of these definitions of energy and momentum:
\begin{equation}
U= \frac{\partial E}{\partial p}\,.
\label{group_vel}
\end{equation}
Our goal is to numerically solve Eqs.~(\ref{eq_caviton}) and (\ref{boundary_cond_2}). However, before addressing this task, it is useful to present some analytical insights.
 In particular, we derive three virial identities and analyze the behavior of the solutions in the limits $E \to 0$ and $E \to \infty$.

\subsection{Virial identities}

By a series of manipulations that consist in using Eq. (\ref{eq_caviton}), the definitions (\ref{pE_defs}) and
integration by parts, one can derive three identities that any solution must satisfy.
This kind of relations are usually called virial identities or Pohozaev identities and in particular they are useful to check the
numerical solutions.
By working with the expression $\int |\nabla (\psi -\psi_{cr})|^2 dxdy$, we can derive:
\begin{eqnarray}
E&=&p\,U - \int \Big[\frac14 |\psi|^4 \big(2 \ln(|\psi|^2) + 1\big) + \nonumber\\
&-& \psi_{cr} {\rm Re(\psi)} \Big(|\psi|^2\ln (|\psi|^2|) + \frac{1}{2\sqrt{e}}\Big)  \Big] dxdy\,.
\label{virial1}
\end{eqnarray}
The trivial identity $(x \partial_x \psi^*) (i U \partial_x \psi) 
+(x \partial_x \psi) (-i U \partial_x \psi^*) =0$ yields, after some computations:
\begin{equation}
E= 2 \int |\partial_x \psi|^2 dx dy\,.
\label{virial2}
\end{equation}
Finally, by integrating $(y \partial_y \psi^*) (i U \partial_x \psi) 
+(y \partial_y \psi) (-i U \partial_x \psi^*) $ in two different ways, we get:
\begin{equation}
U\,p= \int \left[|\psi|^4 \left(\log (|\psi|^2) - \frac12 \right) + \frac{1}{\sqrt{e}}|\psi|^2  \right] dxdy \,.
\label{virial3}
\end{equation}

\subsection{Transonic limit ($U\to U_0$, $E \to 0$)}

Eqs. (\ref{virial1})-(\ref{virial3}) are consistent with the trivial limit in which there is no traveling
wave ($E=p=0$, $\psi=\psi_{cr}$). 
 We now look for solutions of small energy which are close to this limit.
It turns out
 that solutions with infinitesimal $E$ and $p$ are transonic, meaning that they are close but below the speed of
sound ($U \to U_0^-$), with $U_0$ given in Eq. (\ref{sound_speed}). 
An approximate analytic solution in this transonic limit can be found as a small perturbation from the plane wave.
Introducing an infinitesimal $\epsilon \to 0$  parameter, the wavefunction and velocity can be expanded as:
\begin{eqnarray}
f &=&  \psi_{cr} + \epsilon^2 f_1 + \epsilon^4 f_2 + \dots \nonumber\\
g &=& \epsilon g_1 + \epsilon^3 g_2 + \dots \nonumber\\
U &=& U_0 + \epsilon^2 U_1 + \epsilon^4 U_2\,,
\end{eqnarray}
with $f$, $g$ the real and imaginary parts of $\psi$, namely $\psi= f+ i g$.
Rescaling coordinates:
 \begin{equation}
 \hat x = \epsilon \, x \,,\qquad \hat y = \epsilon^2 y\,,
 \end{equation}
 and expanding Eq. (\ref{eq_caviton}), we can solve for the leading terms in the expansion.
 The absolute value of $U_1$ is arbitrary and can be reabsorbed into $\epsilon$, but 
 its sign has to be negative to have solutions.
 Fixing $U_1=- \frac12 e^{-\frac14}$, we get after lengthy but straightforward computations:
 \begin{equation}
 g_1= \frac{-12\hat x}{5(\hat x^2 + \hat y^2 + 3\sqrt{e} )}\,,\qquad f_1 = - \frac{12 e^\frac14 (\hat x^2 +5 \hat y^2 + 15\sqrt{e} )}{25
 (\hat x^2 + \hat y^2 + 3\sqrt{e} )^2}\,.
 \label{transonic_sol}
\end{equation}
At leading order in $\epsilon$, we have:
\begin{equation}
E\approx p\, U_0 
\label{EpU0}
\end{equation}
It is worth mentioning that the equations and solutions that appear in this transonic limit are a version of what is found in the 
Kadomtsev-Petviashvili model, cf. \cite{manakov1977,kuznetsov1995}. 

\subsection{Vortex and antivortex ($U\to 0$, $E\to \infty$)}

Suppose that we have a vortex and an antivortex with $l=\pm 1$ embedded in the critical plane wave. Since the total
vorticity is zero, it can be compatible with the boundary condition (\ref{boundary_cond}). 
Moreover, if they are separated by a large distance, their profiles will be only mildly affected by each other.
Thus, let us consider a vortex at $(x,y)=(0,L)$ and an antivortex at $(0,-L)$ and postulate the following
ansatz:
\begin{equation}
\psi =  \psi_{cr}^{-1} \psi_{1}(r_1) \psi_{1}(r_2) e^{i(\theta_1 - \theta_2)}\,,
\label{v_antiv_ansatz}
\end{equation}
where $\psi_{1}$ is the profile for $l=1$ (see Eq. (\ref{psi_vort})) and $(r_1,\theta_1)$, $(r_2,\theta_2)$ are polar coordinates
centered at the vortex and the antivortex, respectively.
This expression can provide a good approximation to an actual solution when the distance between vortex and antivortex is
much larger than the size of each vortex core, say $L  \gg R_v$, where
\begin{equation}
R_v \approx 5.1 \qquad (|l|=1)\,,
\label{Rvl1}
\end{equation}
 is the half-width at half-maximum of $|\psi_1|^2$, as can be found by computing the numerical solution for 
 $\psi_1(r)$ in (\ref{psi_vort}).
  
In the convention of Eq. (\ref{eqadim}),
the velocity induced by an $l=1$ point-like vortex on another singularity is $2/R$ where $R$ denotes their mutual distance 
(see {\it e.g} \cite{paredes2023polygons} and references therein). Therefore, the velocity of the vortex-antivortex configuration
will be:
\begin{equation}
U \approx \frac{1}{L} ,
\label{ULval}
\end{equation}
which, obviously, tends to 0 in the $L\to\infty$ limit.
Using the virial identity (\ref{virial3}), we can find the values of $p$ and $E$ in this limit.
From the numerical profile $\psi_1(r)$, we can compute the integral in Eq. (\ref{virial3}) for a single vortex:
\begin{equation}
2\pi \int_0^\infty  r \left[\psi_1^4 \left(\log (\psi_1^2 - \frac12 \right) + \frac{1}{\sqrt{e}}\psi_1^2  \right] dr \approx 3.808\,.
\end{equation}
If the vortex and antivortex are far apart from each other, the integrand in (\ref{virial3}) has support only around each singularity and
the total integral is simply twice the integral around each of them. Namely $pU\approx 2 \times 3.808$ and
\begin{equation}
p \approx \frac{7.616}{U}\,.
\label{pUval}
\end{equation}
Using (\ref{group_vel}), we see that the energy diverges logarithmically:
\begin{equation}
E  \approx  7.616 \ln p +  const\,.
\label{EplargeU}
\end{equation}
The value of the additive constant has been computed numerically, see subsection \ref{subsec:num}.

\subsection{Numerical solutions ($0<U<U_0$)}
\label{subsec:num}

The next step is to find numerical solutions of (\ref{eq_caviton}), (\ref{boundary_cond_2}).  
We will follow the method devised in \cite{chiron2016travelling}, which consists in
working with the following heat flow equation, which depends on a real parameter $\alpha$:
\begin{equation}
\frac{\partial u}{\partial \tau} = -i(\alpha-p(u)) \partial_x u + \nabla^2 u  - \left(|u|^2 \ln (|u|^2) + \frac{1}{2\sqrt{e}}\right) u\,,
\label{eq_heat_flow}
\end{equation}
where $u(\tau,x,y)$ is a complex function and $p(u)$ is the momentum as defined in Eq. (\ref{pE_defs}), computed with the function $u$,
and $\tau$ is an auxiliary, imaginary time parameter.
The parameter $\alpha$ is a constant that we fix for each computation and defines the solution we are looking for.
 If after $\tau$-evolution the function $u$ converges such
that $\frac{\partial u}{\partial \tau}=0$, we then have
\begin{equation}
\psi (x,y)=\lim_{\tau \to \infty} u(\tau, x,y)\,,
\end{equation}
 a solution of Eq. (\ref{eq_caviton})
with:
\begin{equation}
U = \alpha - p(u)\,.
\label{Ualphap}
\end{equation}
In order to check convergence, we define the error of the approximation:
\begin{equation}
{\cal E} (u) = \frac{\lVert  -i(\alpha-p(u)) \partial_x u + \nabla^2 u  - \left(|u|^2 \ln (|u|^2) + \frac{1}{2\sqrt{e}}\right) u  \rVert}{
\lVert  (\alpha-p(u)) \partial_x u \rVert}\,,
\end{equation}
where $\lVert f \rVert  = \int |f|^2 dxdy$. 
We check that the evolution tends to reduce this error and define a tolerance 
limit where to stop the computation. 
Convergence is achieved if we start with an ansatz for $u$ sufficiently close to the final solution for the $\alpha$ at hand.
Therefore, we can compute a first solution using the vortex-antivortex ansatz of Eq. (\ref{v_antiv_ansatz})
as a starting point for the heat flow. For a  
vortex-antivortex distance $L$ satisfying $L \gg R_v$, we have:
\begin{equation}
\alpha = U + p \approx U+ \frac{7.616}{U} \approx \frac{7.616}{U}
\approx 7.616 L\,,
\end{equation}
where we have used (\ref{Ualphap}), (\ref{pUval}) and (\ref{ULval}). 
Once we have this first solution, we can find the rest of the family by iteratively reducing $\alpha$
in small steps. At each step, we use the previous solution
as the initial condition of (\ref{eq_heat_flow}) and 
 look for convergence of the heat flow with the next $\alpha$ \cite{chiron2016travelling}.
 
 With this method, we have found the full family of solutions $U \in (0,U_0)$ connecting the transonic limit of Eq. (\ref{v_antiv_ansatz})
 to the $E \to 0$ limit of Eq. (\ref{transonic_sol}).
 For each value of $\alpha$ in the range $\infty > \alpha > U_0$, there exists a unique solution within the family. As $\alpha$ decreases, the momentum $p$ decreases monotonically, while the velocity $U$ increases monotonically. The distance between phase singularities also decreases monotonically with decreasing $\alpha$ until it reaches zero at a certain $\alpha$, beyond which no phase singularities exist.
 
\begin{figure}[h!]
\begin{center}
\includegraphics[width=\columnwidth]{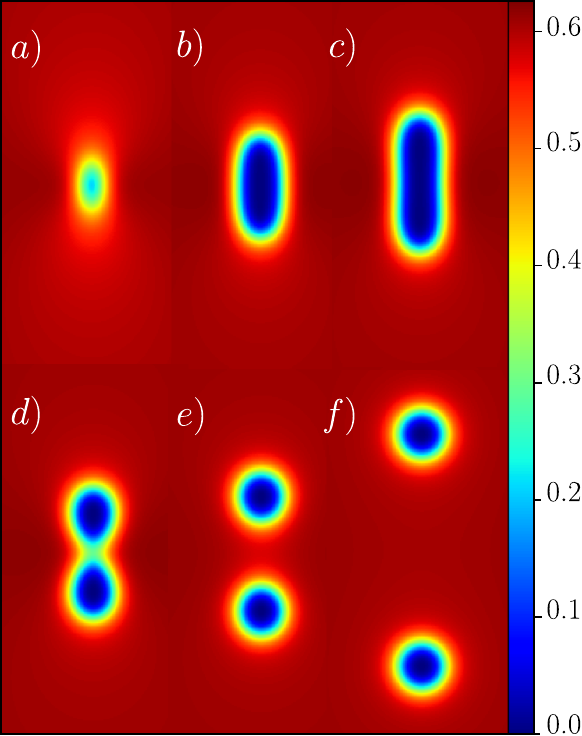}
\end{center}
\caption{Some bubble profiles that solve Eqs. (\ref{eq_caviton}), (\ref{boundary_cond_2}). We depict $|\psi(x,y)|^2$ for
solutions with: a) $p  = 5.7$, b) $p=27  $, c) $p=45$, d) $p=57 $, e) $p=85$, f) $p= 185$. Each of the six windows spans the spatial range 
$x \in (-20, 20)$, $y\in (-40,40)$. }
\label{fig2}
\end{figure}

We depict some results in Fig. \ref{fig2}. Remember that we have assumed right-ward velocities along the $x$-axis. Obviously, for velocities along different directions the profiles should be
rotated.

%%%%%%%%%%%%%%%%%%%%%% FIGURE 2 %%%%%%%%%%%%%%
%%%%%%%%%%%%%%%%%%%%%%%%%%%%%%%%%%%%%%%%%%

\begin{figure}[h!]
\begin{center}
\includegraphics[width=\columnwidth]{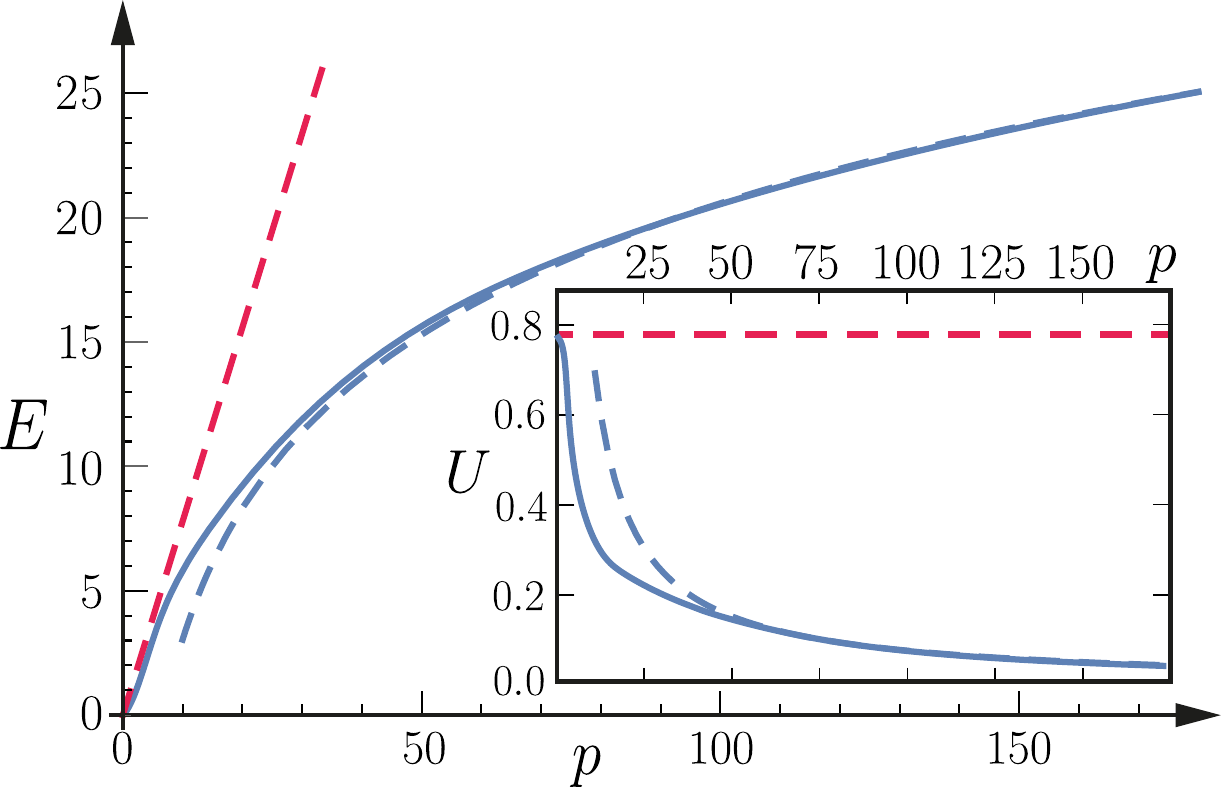}
\end{center}
\caption{Numerical results for the dispersion relation $E(p)$ of the stable bubble solutions.  The red dashed line
corresponds to $E=U_0 p$, the asymptotic behaviour for $p\to 0$. The blue dashed line is
$E= 7.616\log p - 14.5$, valid in the limit of large $p$, see Eq. (\ref{EplargeU}). 
In the inset, we depict $U(p)=\frac{\partial E}{\partial p}$, compared to its asymptotic expressions.
}
\label{fig3}
\end{figure}

In Fig. \ref{fig3}, we depict the energy and the velocity as functions of momentum, and compare the numerical results
to the asymptotic expressions, Eqs. (\ref{EpU0}), (\ref{EplargeU}), (\ref{pUval}). The graph shows how the numerical results smoothly interpolate
between the analytical approximations for transonic limit and the limit of well-separated vortex-antivortex pairs.

%%%%%%%%%%%%%%%%%%%%%% FIGURE 3 %%%%%%%%%%%%%%
%%%%%%%%%%%%%%%%%%%%%%%%%%%%%%%%%%%%%%%%%%

\begin{figure}[h!]
\begin{center}
\includegraphics[width=\columnwidth]{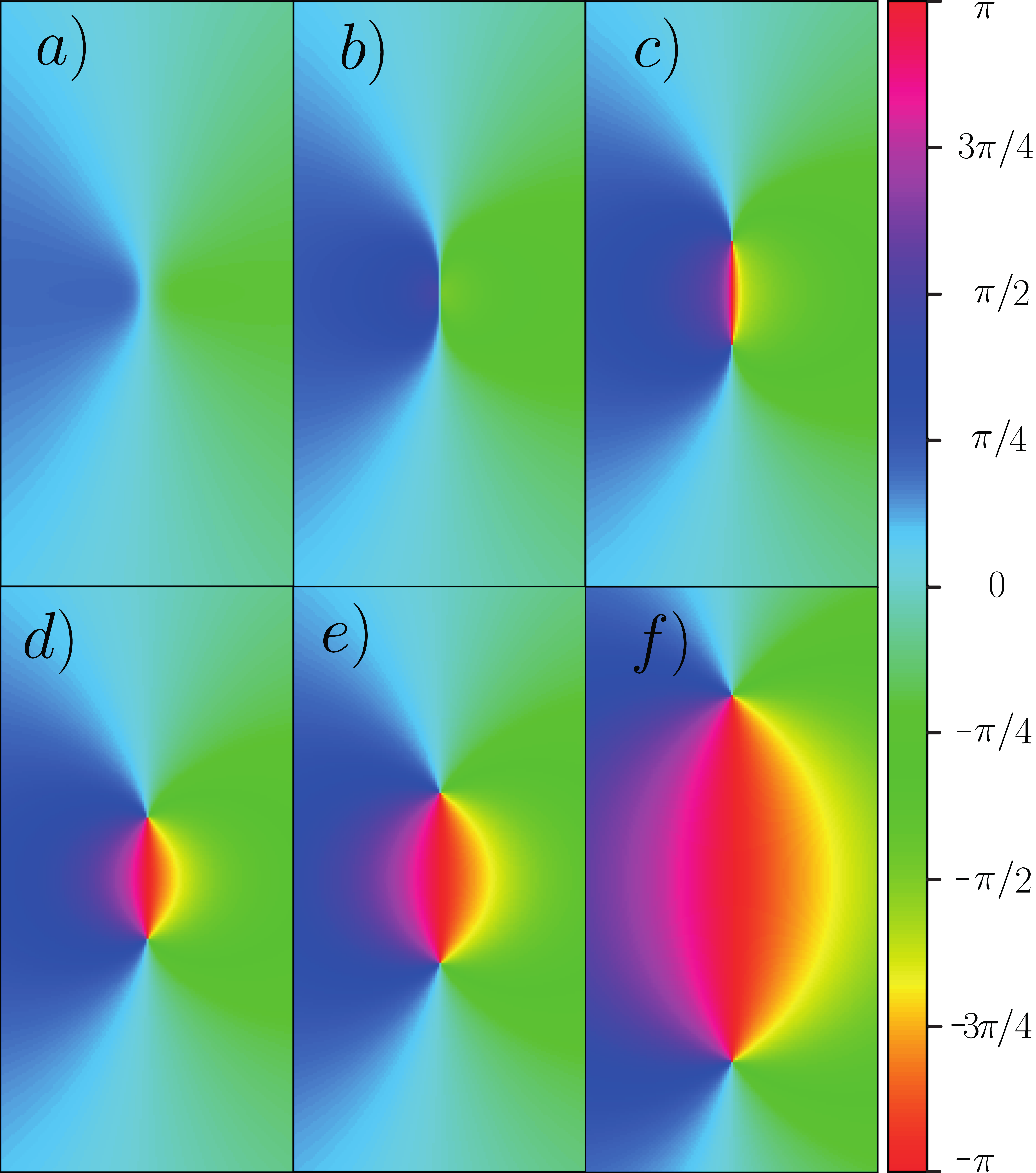}
\end{center}
\caption{The phase of the wavefunction $\arg(\psi(x,y))$ for the six cases depicted in Fig. \ref{fig1}. In panels c)-f), two phase singularities can be clearly appreciated. 
In panels a) and b) there is a non-trivial phase profile, but there are no phase singularities.}
\label{fig3b}
\end{figure}

From the graphs of Fig. \ref{fig2}, we can readily see that there are three qualitatively distinct cases. For small velocities,
$0<U<U_{sc}$ ($p>p_{sc}$) we have a vortex and antivortex pair with separate vortex cores. Namely, there are two distinct dips
embedded within the critical density fluid. At 
$U_{sc}$, both cores get joint forming a ``single core'' and there is a region $U_{sc} < U < U_{rp}$ 
($p_{sc} > p > p_{rp}$)
for which there are still two phase singularities, but they are embedded in a single dip of the density.
As speed grows, both singularities get closer to each other and, eventually, at $U=U_{rp}$, they
meet so for $U_{rp} < U < U_0$ ($p_{rp} > p  > 0$) there is no phase singularity and we have what is usually called a 
rarefaction pulse. Rarefaction pulses where first discussed in \cite{jones1982motions}. They are travelling waves with a dip in the density and a non-trivial phase
profile but without phase singularities. To the best of our knowledge, they have not been discussed previously in the context of QDs.
The existence or not of phase singularities can also be visualized by depicting the phase of $\psi$, see Fig. \ref{fig3b}.

%%%%%%%%%%%%%%%%%%%%%% FIGURE 3b %%%%%%%%%%%%%%
%%%%%%%%%%%%%%%%%%%%%%%%%%%%%%%%%%%%%%%%%%

\begin{figure}[h!]
\begin{center}
\includegraphics[width=\columnwidth]{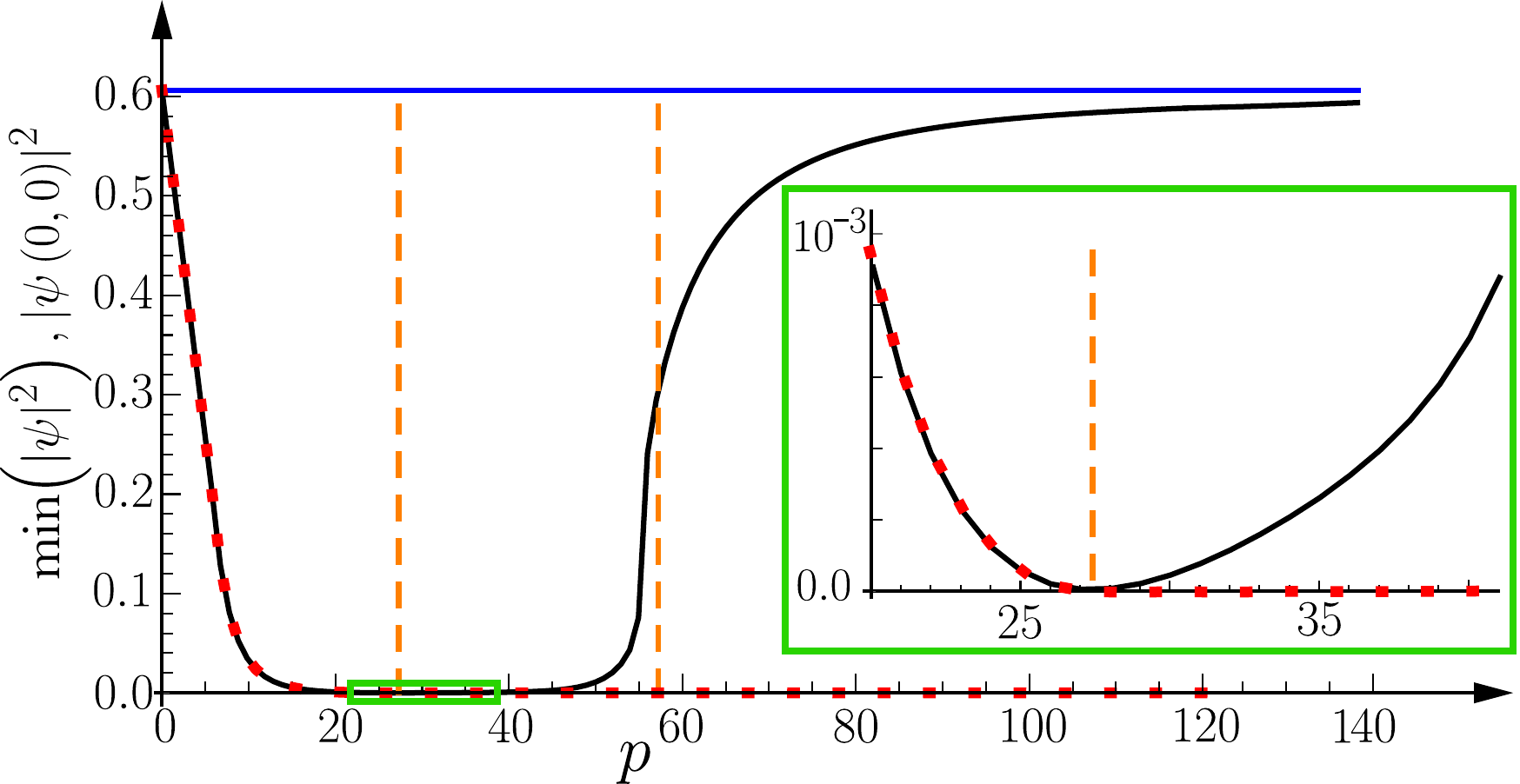}
\end{center}
\caption{The minimum of the density ${\rm min}[|\psi(x,y)|^2]$ (red dashed line)  and the density at the origin $|\psi(0,0)|^2$ (black solid line) 
as a function of $p$
for the family of bubble eigenstates, including vortex-antivortex pairs.
The vertical dashed lines mark the transition points $p=p_{rp}\approx 27$ and $p=p_{sc}\approx 57$, respectively. The horizontal blue solid line
is $\psi_{cr}^2$. In the inset, we enlarge the region near $p=p_{rp}$. For $p<p_{rp}$, ${\rm min}[|\psi|^2]$ and  $|\psi(0,0)|^2$ are equal and non-vanishing.
For $p_{rp}<p$, there are phase singularities implying that ${\rm min}[|\psi|^2]$ is identically zero and, on the other hand, $|\psi(0,0)|^2$ starts growing.
}
\label{fig4}
\end{figure}

These regimes, analogous to those found in the seminal paper \cite{jones1982motions},
can be explained in terms of the quantities depicted in Fig. \ref{fig4},
namely the minimum value of $|\psi(x,y)|^2$ and the value of $|\psi(0,0)|^2$, the density at the coordinate origin.
For $p<p_{rp}$ (Fig. \ref{fig2} a)), ${\rm min}[|\psi|^2]$ is non-vanishing and this minimum is
located at the center, namely ${\rm min}[|\psi|^2]=|\psi(0,0)|^2 \neq 0 $ and both lines of Fig. \ref{fig4} overlap.
The transition point $p = p_{rp} \approx 27$ (Fig. \ref{fig2} b)) is defined by ${\rm min}[|\psi|^2]=|\psi(0,0)|^2 = 0 $.
For $p_{rp}< p<p_{sc}$  (Fig. \ref{fig2} c)), there are two phase singularities where $|\psi|$ has to vanish
and, therefore, ${\rm min}[|\psi|^2]=0$. Since the singularities lie within the same density dip, the central 
point has small density, $0< |\psi(0,0)|^2 \ll \psi_{cr}^2$. For $p_{sc} < p$ (Figs. \ref{fig2} e), f)), there are two separate density valleys and the
density in between them is close to the critical one, namely $|\psi(0,0)|^2$ approaches $\psi_{cr}^2$.
The transition between the cases of two separate cores and a single dip is gradual
and therefore $p_{sc}$ is not sharply defined. We can provide a value by defining it as the point where
$|\psi(0,0)|^2 = \psi_{cr}^2/2$, from which we get $p_{sc} \approx 57$ (Fig. \ref{fig2} d)).

%%%%%%%%%%%%%%%%%%%%%% FIGURE 4 %%%%%%%%%%%%%%
%%%%%%%%%%%%%%%%%%%%%%%%%%%%%%%%%%%%%%%%%%

The numerical value of \(p_{sc}\) can be understood heuristically as follows: the quantum droplet can be viewed as a liquid with surface tension along the domain
 walls that connect regions of low density to regions of critical density \cite{novoa2009pressure,paredes2024variational}. The transition to a single core occurs when
  the surface area enclosing the single core becomes smaller than the combined surface area surrounding two separate cores. Thus, the transition to a single core is characterized
  by a vortex-antivortex distance $2L=2L_{sc}$ with:
\begin{equation}
2 \pi R_v + 4 L_{sc} \approx 4 \pi R_v \,,
\label{approx_transition}
\end{equation}
where we have approximated the single core boundary by two semicircles of radius $R_v$ connected by straight lines. 
From here, using (\ref{ULval}), (\ref{pUval}) and (\ref{Rvl1}), we find $p_{sc} \approx 61$ which is a fairly reasonable
approximation to the value quoted above, that was obtained from the numerical solutions.

\begin{figure}[h!]
\begin{center}
\includegraphics[width=\columnwidth]{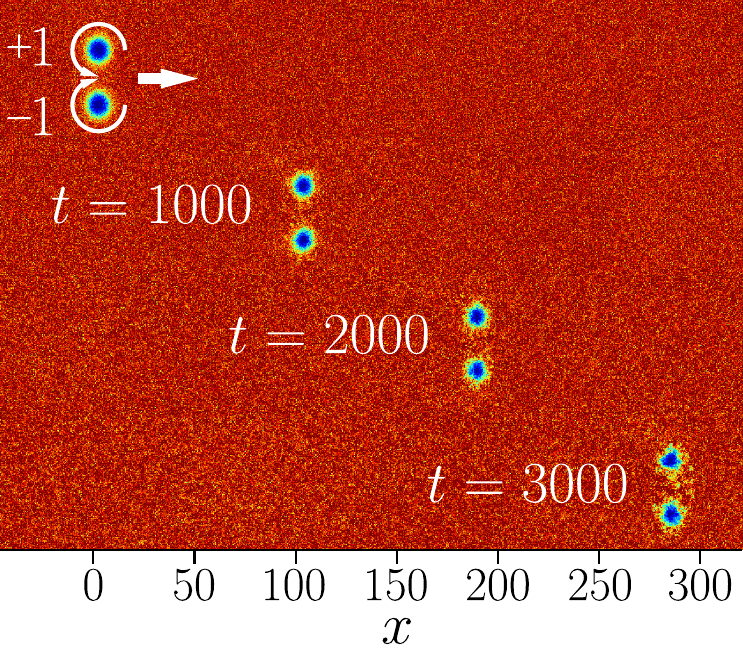}
\end{center}
\caption{Real-time evolution of the eigenstate shown in Fig. \ref{fig2}.b). The initial wavefunction includes a 
noise perturbation of 8\% relative to $\psi_{cr}$. We plot the result of the simulation at different values of the adimensional time $t$. 
In the first image, we highlight with arrows the phase gradient due to the phase singularities.
Despite the noisy background, the vortex-antivortex pair propagates
with constant velocity without any serious distortion to its density profile. }
\label{fig4b}
\end{figure}

\begin{figure}[h!]
\begin{center}
\includegraphics[width=\columnwidth]{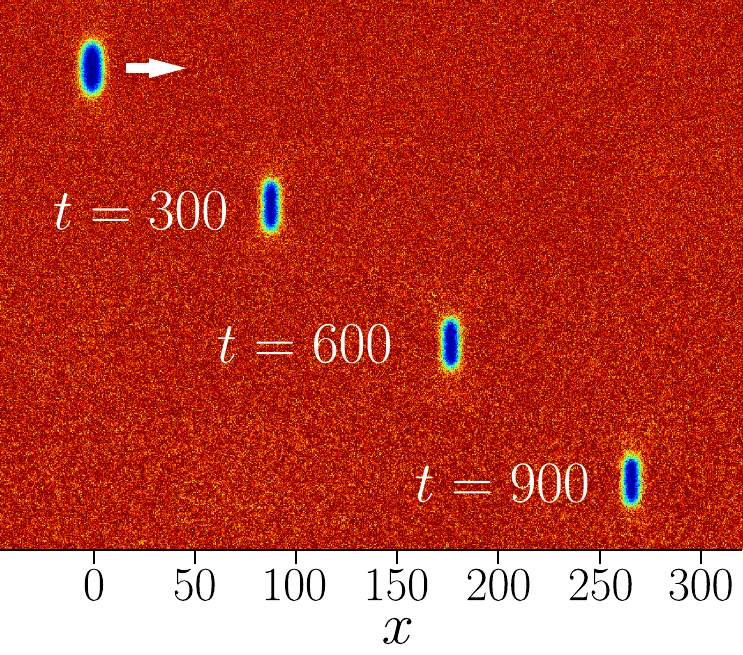}
\end{center}
\caption{Real-time evolution of the eigenstate shown Fig. \ref{fig2}.e). The initial wavefunction includes a 
noise perturbation of 8\% relative to $\psi_{cr}$. Despite the noisy background, the bubble propagates mostly unchanged. Note the different time scales with respect
to the case of Fig, \ref{fig4b},
highlighting that vortex-antivortex pairs move at lower velocities compared to rarefaction pulses.  }
\label{fig4c}
\end{figure}

Finally, it is worth highlighting the behavior of the eigenstates discussed in this section when propagated in real time. Using the numerically obtained
 solutions as initial conditions for (\ref{eqadim}) and (\ref{boundary_cond}), we have verified that the bubbles indeed travel at a constant speed while
  maintaining their shape. This confirms the stability of the entire family of solutions in this regard, which is consistent with the fact that these solutions correspond 
  to the minimal energy configuration for a given momentum. 
As numerical evidence for this stability,  in Figs. \ref{fig4b} and \ref{fig4c} we depict two examples of the evolution of bubbles that maintain their shape, even introducing noisy initial conditions.
  A more detailed verification of stability could be performed by adapting the methods presented in \cite{berloff2004motions} 
  to this specific context; however, such an analysis lies beyond the scope of this work.

%%%%%%%%%%%%%%%%%%%%%% FIGURE 4b %%%%%%%%%%%%%%
%%%%%%%%%%%%%%%%%%%%%%%%%%%%%%%%%%%%%%%%%%

%%%%%%%%%%%%%%%%%%%%%% FIGURE 4b %%%%%%%%%%%%%%
%%%%%%%%%%%%%%%%%%%%%%%%%%%%%%%%%%%%%%%%%%

\section{Vortex-antivortex pairs with higher topological charge}
\label{sec:largel}

In the two-dimensional QD model of Eq. (\ref{eqadim}), there exist stable vortices with topological charges larger than one \cite{li2018two,salgueiro2024stability}.
Thus, it is natural to wonder whether they can be used as building blocks to construct other families of shape-preserving traveling bubbles, apart from the
one constructed above starting from singly charged vortices. We address here this question by adapting the procedure of section \ref{sec:stable}.
First we consider the case of a well-separated vortex-antivortex pair with $|l|>1$, generalizing Eq. (\ref{v_antiv_ansatz}) to:
\begin{equation}
\psi =  \psi_{cr}^{-1} \psi_{|l|}(r_1) \psi_{|l|}(r_2) e^{il(\theta_1 - \theta_2)}\,.
\label{v_antiv_ansatz_2}
\end{equation}
The $\psi_{|l|}(r)$ functions can be found numerically, but it is possible to write an analytic approximation, that is very
accurate for $|l|>1$, namely \cite{paredes2024variational}:
\begin{equation}
\psi_{|l|}(r) \approx \frac{\psi_{cr}}{\sqrt2} \sqrt{1+ \tanh (\kappa\,r - 2 l^2)}\,,
\label{psilprofile}
\end{equation} 
with $\kappa = \frac{\sqrt{12-\pi^2}}{\sqrt6\,e^\frac14 } \approx 0.4641$. 
With this ansatz, we can readily compute numerically the integral in Eq. (\ref{virial3}) and find large momentum asymptotic approximations for
$U(p)$, $E(p)$ for separate vortex-antivortex pairs. In the $l=2$ case, we find:
\begin{equation}
U\approx \frac{28.11}{p}\,, \quad E= 28.11 \log p + const\,.
\label{UEpl2}
\end{equation}
On the other hand, for a vortex-antivortex separation of $2L$, we have $U \approx l/L$ and, thus, for $l=2$, we get:
\begin{equation}
p \approx \alpha \approx 14.06 L \,.
\label{pLl2}
\end{equation}
We insert this value along with the ansatz into the heat flow of 
Eq. (\ref{eq_heat_flow}) in order to find a first solution. 
Then, again,  we reduce the value of $\alpha$ step by step, gradually decreasing the separation between the phase singularities, and find a family of solutions.
For large $p$, we get separate vortex and antivortex. For intermediate $p$, around $p_{sc} \approx 420$, the two cores merge into a single dip. 
As in the case of section \ref{sec:stable}, a rough approximation to this value can be found from Eq. (\ref{approx_transition}). Using Eqs. 
(\ref{pLl2}) and $R_v = 2 l^2/\kappa \approx 17.2  $ (see (\ref{psilprofile})), we obtain $p_{sc} \approx 381$ which is within 10\% of the actual numerical value.

\begin{figure}[h!]
\begin{center}
\includegraphics[width=\columnwidth]{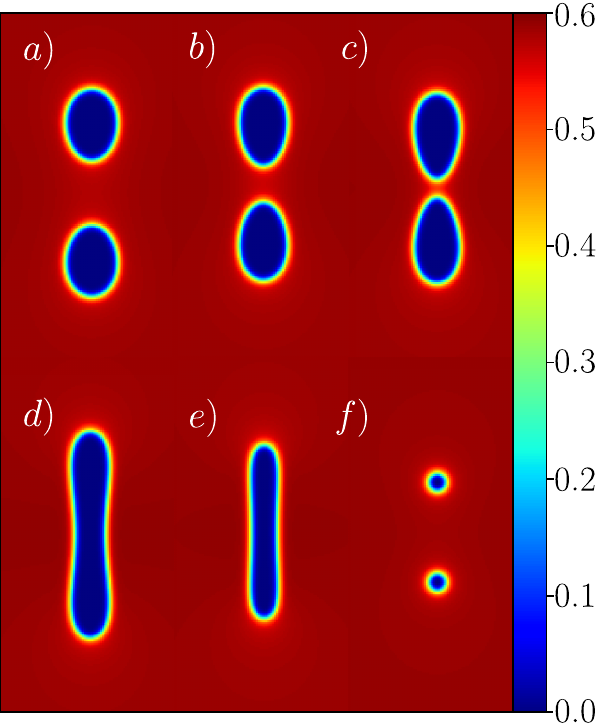}
\end{center}
\caption{Some bubble profiles of the family starting with an $l=\pm 2$ vortex-antivortex pair. We depict $|\psi(x,y)|^2$ for
solutions with: a) $p  = 561 $, b) $p= 471 $, c) $p=431$, d) $p=386 $, e) $p=266$, f) $p= 211$. Each of the six windows spans the spatial range 
$x \in (-50, 50)$, $y\in (-100,100)$. Notice that the case depicted in f) corresponds to $l=\pm 1$ vortex-antivortex, namely the heat flow
has converged to a solution of the family discussed in section \ref{sec:stable}.}
\label{fig5}
\end{figure}

As we continue reducing $\alpha$,
eventually, for smaller $p \approx 250$, the single dip gets
split again, and the method simply converges to the separate vortex and antivortex with $l=\pm 1$, namely the case studied in section \ref{sec:stable}.
Some examples are presented in Fig. (\ref{fig5}).

%%%%%%%%%%%%%%%%%%%%%% FIGURE 5 %%%%%%%%%%%%%%
%%%%%%%%%%%%%%%%%%%%%%%%%%%%%%%%%%%%%%%%%%

\begin{figure}[h!]
\begin{center}
\includegraphics[width=\columnwidth]{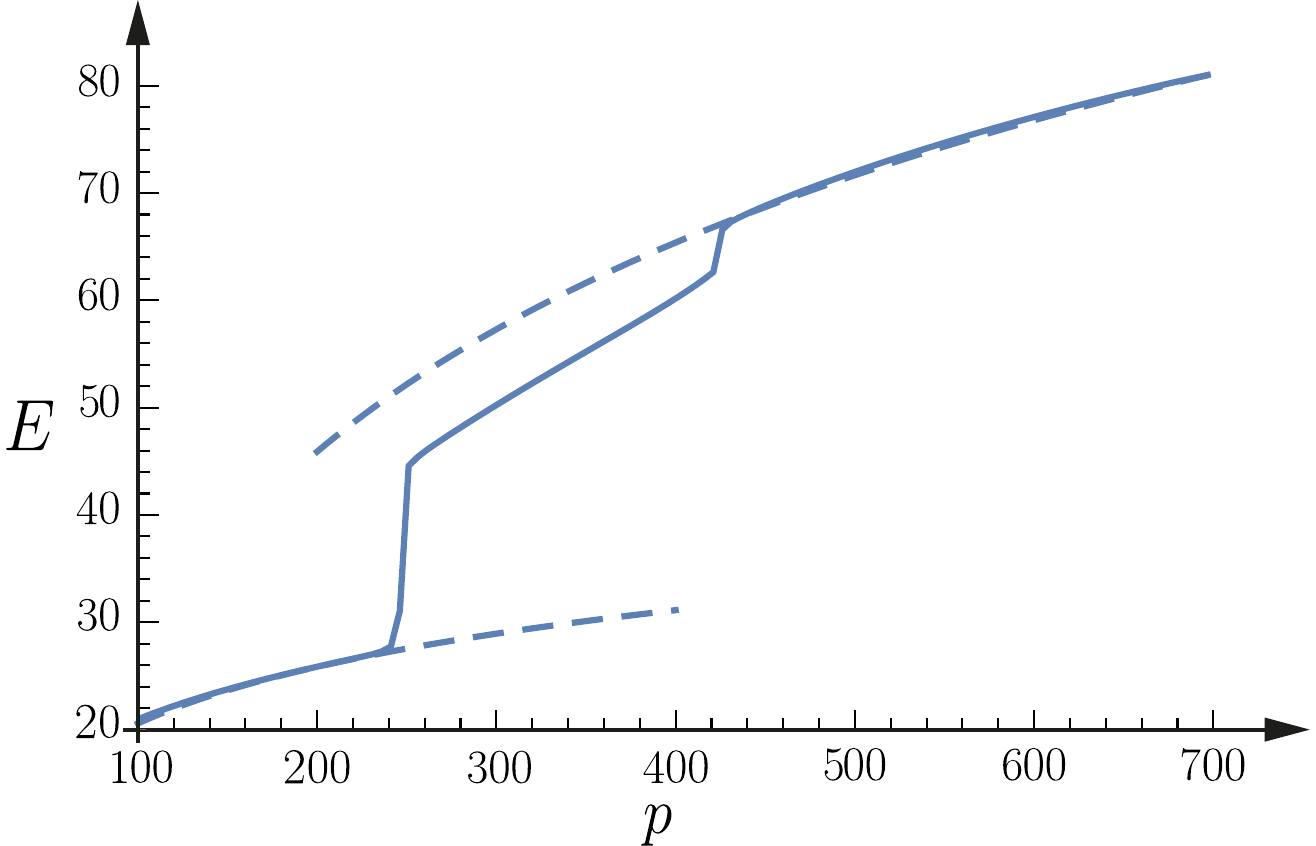}
\end{center}
\caption{Numerical results for the dispersion relation $E(p)$ of the bubble solutions of the family starting with an $l=\pm 2$ vortex-antivortex pair.
  The upper dashed lines corresponds to Eq. (\ref{UEpl2}) with $const=-103$, the $p\to \infty$ asymptotic behaviour for well separated $l=\pm 2$
  vortex and antivortex. The lower dashed line is Eq. (\ref{EplargeU}) corresponds to well separated $l=\pm 1$ vortex and antivortex,
  see panel f) of Fig. \ref{fig5}. 
}
\label{fig6}
\end{figure}

\begin{figure}[h!]
\begin{center}
\includegraphics[width=\columnwidth]{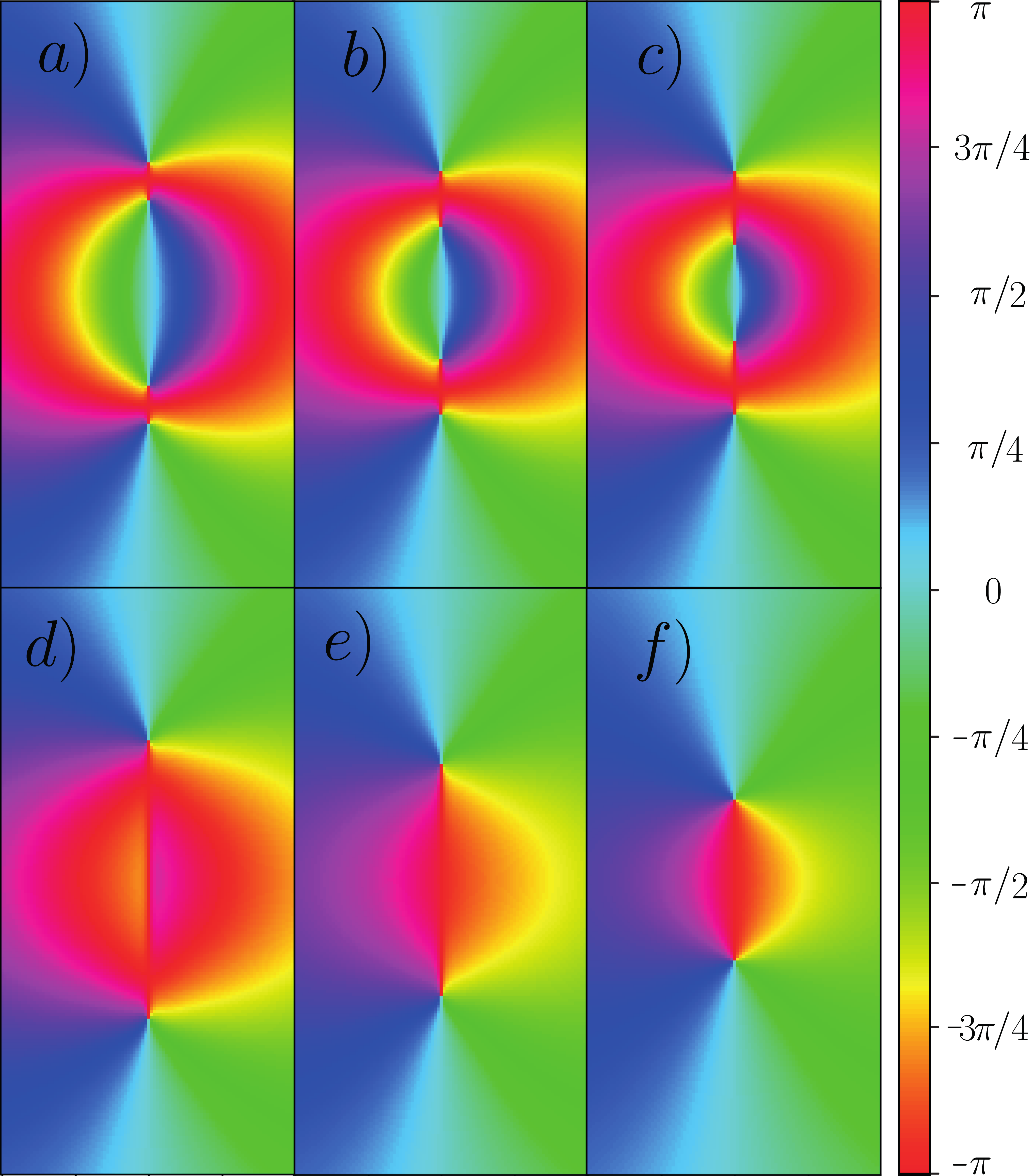}
\end{center}
\caption{The phase of the wavefunction $\arg(\psi(x,y))$ for the six cases depicted in Fig. \ref{fig5}.}
\label{fig5b}
\end{figure}

The dispersion relation $E(p)$ for the family built from $l=\pm 2$ vortices is presented in Fig. (\ref{fig6}). 
In particular, notice that in this case there are two sudden jumps at the two transitions between
separated cores and a single dip, and between a single dip and separated cores with unit topological charge.
The graphs show that, unlike the case of section \ref{sec:stable}, there is no smooth interpolation between the large $p$ and small $p$ limits.
In order to understand better the properties of the solutions along the family, we depict in Fig. \ref{fig5b} the phase of the wavefunction.
Notice that the vortices of topological charge two in fact include two separate order one phase singularities within the density dip at the vortex core.
As the vortex and antivortex get closer, the bubbles get elongated and the phase singularities are farther apart from each other. In the transition to a single
density dip, only two phase singularities of order $\pm1$ survive. Eventually, at smaller $p$, the solution with a single trough ceases to exist and we 
fall into the case of well-separated vortices with $l=\pm 1$.

%%%%%%%%%%%%%%%%%%%%%% FIGURE 6 %%%%%%%%%%%%%%
%%%%%%%%%%%%%%%%%%%%%%%%%%%%%%%%%%%%%%%%%%

%%%%%%%%%%%%%%%%%%%%%% FIGURE 6b %%%%%%%%%%%%%%
%%%%%%%%%%%%%%%%%%%%%%%%%%%%%%%%%%%%%%%%%%

\begin{figure}[h!]
\begin{center}
\includegraphics[width=\columnwidth]{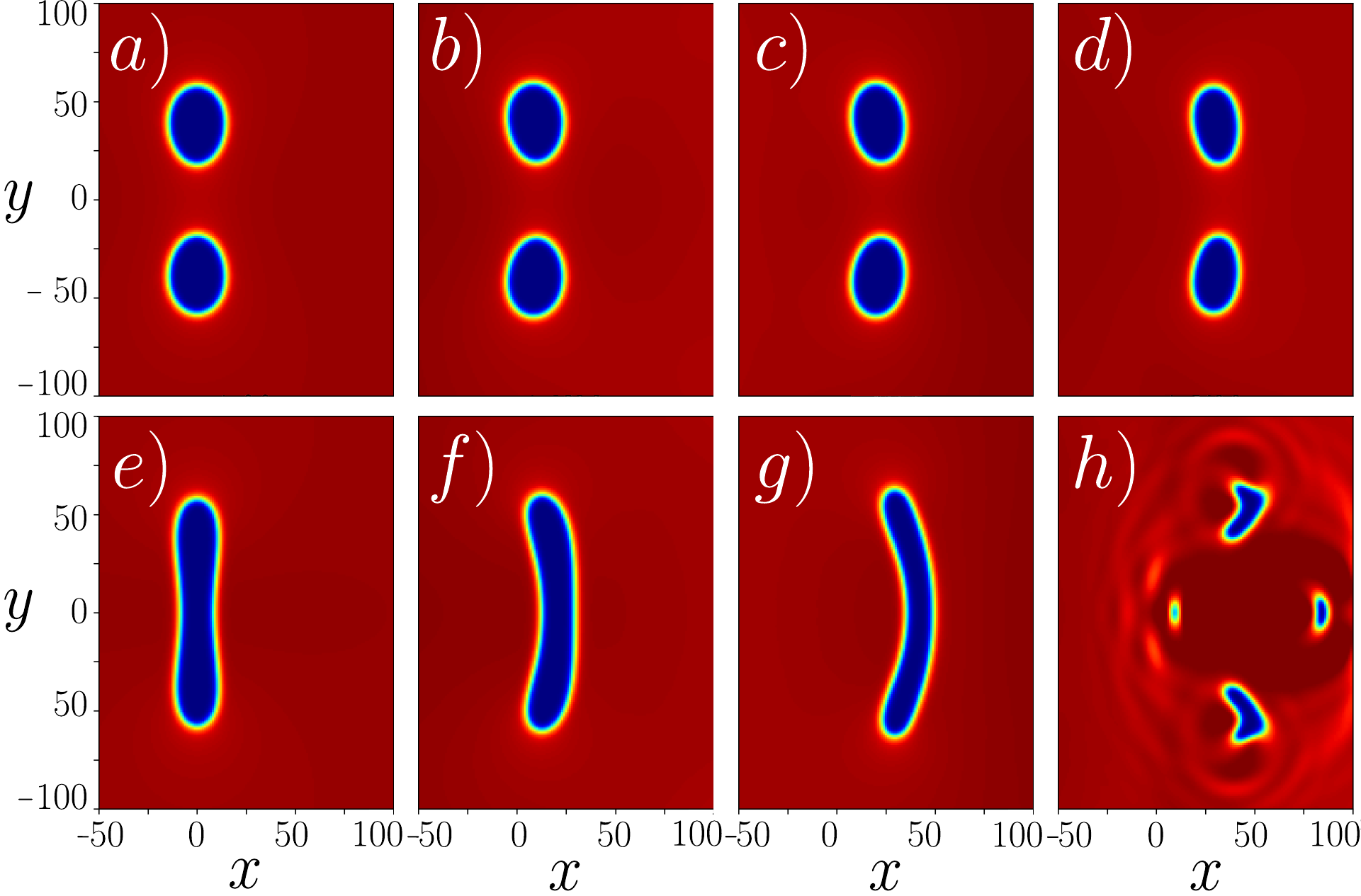}
\end{center}
\caption{Two examples of the simulation of the evolution of unstable bubbles. 
The graphs depict $|\psi(x,y)|^2$ with the same  color code as Figs. \ref{fig2}, \ref{fig5}.
The upper panels correspond to the evolution of the state of frame a) in Fig. \ref{fig5},
at times a) $t=0$, b) $t=180$, c) $t=360$, d) $t=540$.
The lower panels correspond to the evolution of the state of frame d) in Fig. \ref{fig5},
at times e) $t=0$, f) $t=180$, g) $t=360$, h) $t=540$.
This latter case decays, leading to the formation of two density dips with $ l = \pm 1$ phase singularities, accompanied by a forward-moving rarefaction pulse
 and a backward-moving rarefaction pulse, as shown in panel h). These rarefaction pulses are those discussed in section \ref{sec:stable}, although, obviously, in the dynamical process only 
 approximations to the actual eigenstates are formed.
}
\label{fig7}
\end{figure}

Finally, it is crucial to address the stability of these configurations. The family described in this section does not correspond to the global minimum of $E$ for a given $p$, which
 suggests that these solutions may not persist indefinitely and could eventually decay into those discussed in section \ref{sec:stable}. To investigate this, we have performed 
real-time propagation using a numerical method adapted from \cite{figueiras2018open} and references therein. Our analysis reveals that when the vortex
cores are well separated, the configurations can propagate largely unchanged for extended evolution times, exhibiting only minor oscillations in the
size and shape of the bubbles. However, in cases with a single dip, the instability becomes more pronounced: the dark band bends, stretches, and
 eventually breaks apart, producing a singly charged vortex-antivortex pair alongside other fluid perturbations, such as rarefaction pulses. Two examples 
 of this behavior are shown in Fig.~\ref{fig7}. Additionally, we observe an intermediate scenario (not depicted), where two initially separated but closely 
 positioned vortex cores merge over time into a stretched single dip, which subsequently decays.

%%%%%%%%%%%%%%%%%%%%%% FIGURE 7 %%%%%%%%%%%%%%
%%%%%%%%%%%%%%%%%%%%%%%%%%%%%%%%%%%%%%%%%%

It is possible to repeat the analysis with vortices of higher charges $|l|>2$. We find qualitatively similar results, albeit larger charges tend to make
the configurations more unstable.

\section{Cohesively traveling vortex arrays}
\label{sec:linea}

\begin{figure}[h!]
\begin{center}
\includegraphics[width=\columnwidth]{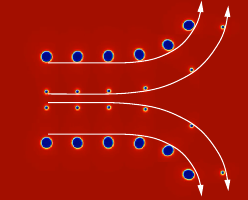}
\end{center}
\caption{Simulation of the evolution of a vortex array with vortices of $l=\pm 1$ and $\pm 2$, with $L_1=25$ and $L_2$ given
in Eq. (\ref{L2L1}). The color code for $|\psi(x,y)|^2$ is the same as that in Figs. \ref{fig2}, \ref{fig5}.
For visualization purposes, we have included the four vortex cores at different values of 
$t$ in the same background. From left to right, the snapshots correspond to $t=0,5000,10000,15000,20000,25000$. The structure moves cohesively with 
the velocity given in Eq. (\ref{UL1val}) up to $t\approx 15000$, after which it gets destabilized. The white arrows are included to 
assist interpretation.
}
\label{fig8}
\end{figure}

In this section, we study one last possibility for nonlinear waves traveling with constant velocity $U$. 
Up to now, we have studied families of solutions that for large $p$ correspond to a vortex-antivortex pair
of $l=\pm 1$ (section \ref{sec:stable}) or $ |l| >1 $ (section \ref{sec:largel}). Here, we consider the possibility of
having more than one vortex-antivortex pair, carefully assembled so they move cohesively, preserving the
form of the vortex array. 
Let us thus suppose that we have $N$ pairs, with pair $i$ having a phase singularity of topological charge $l_i$ placed at
$(0,L_i)$ and a singularity of charge $-l_i$ placed at $(0,-L_i)$. By definition, we take $L_i$ to be positive but the $l_i$ can be positive or negative.
 Due to symmetry, the vortex and antivortex of each pair 
move with the same velocity, and, in the point-like vortex approximation, we can write the velocity for each pair as:
\begin{equation}
v_i = \frac{l_i}{L_i}+ 4 \sum_{j\neq i} \frac{l_j L_j}{L_j^2 - L_i^2}\,,\qquad i,j = 1, \dots, N\, .
\label{vipuntual}
\end{equation}
This expression is readily derived using the fact that the velocity that a point vortex of charge $l$ induces on any other point vortex is $|v| =2 |l|/R$, where $R$ is the distance 
between them. If we want the array of vortices to move cohesively, we need that all the velocities are equal:
\begin{equation}
v_1=  v_2 = \dots = v_N \,.
\label{equalv}
\end{equation}
which is a set of $N-1$ equations. If we fix the topological charges, there are $N$ unknowns $L_i$, which can be easily reduced
to $N-1$ since only the quotients of distances appear, namely $L_i/L_1$ for $i=2,\dots, N$. 
Let us consider the simplest case, $N=2$, $l_1=-1$, $l_2=2$, in order to  illustrate this kind of configurations.
 With these parameters, Eqs. (\ref{equalv}) are reduced to a cubic algebraic equation for $\frac{L_2}{L_1}$:
\begin{equation}
\left(\frac{L_2}{L_1}\right)^3 -6 \left(\frac{L_2}{L_1}\right)^2 + 3  \frac{L_2}{L_1} -2 = 0\,,
\end{equation}
which only has one real solution:
\begin{equation}
\frac{L_2}{L_1} = 2 +3^\frac13 + 3^\frac23 \approx 5.522\,.
\label{L2L1}
\end{equation}
We can compute the value of $U=v_i$ from Eq. (\ref{equalv}):
\begin{equation}
U = \frac{2(3^\frac13 - 3^{-\frac13}) - 1}{L_1} \approx \frac{0.4978}{L_1}\,.
\label{UL1val}
\end{equation}
Having two singularities of order one and two of order two, the asymptotic value of $Up$ for large separations is given the sum of 
those in Eqs. (\ref{pUval}) and (\ref{UEpl2}), namely $\lim_{p\to \infty} Up \approx   35.73$.
Then:
\begin{equation}
\alpha = p + U \approx p \approx \frac{35.73}{U} \approx 71.8 L_1\,,
\label{mu_Array}
\end{equation}
and
\begin{equation}
E \approx 35.73 \log p + const\,.
\end{equation}
The solutions are found to be unstable: while the vortex array can travel cohesively for some time, the structure eventually disintegrates. 
The instability leads to two qualitatively distinct outcomes. In the first scenario, the $l=2$ vortex and the $l=-1$ antivortex located in the $y>0$ region pair 
up and begin moving collectively with a nonzero velocity component in the $y$-direction ($v_y \neq 0$), while simultaneously orbiting around each other. 
A similar process occurs for their mirror counterparts in the $y<0$ half-plane. Fig.~\ref{fig8}
depicts an illustration of this behaviour.

%%%%%%%%%%%%%%%%%%%%%% FIGURE 8 %%%%%%%%%%%%%%
%%%%%%%%%%%%%%%%%%%%%%%%%%%%%%%%%%%%%%%%%%

\begin{figure}[h!]
\begin{center}
\includegraphics[width=\columnwidth]{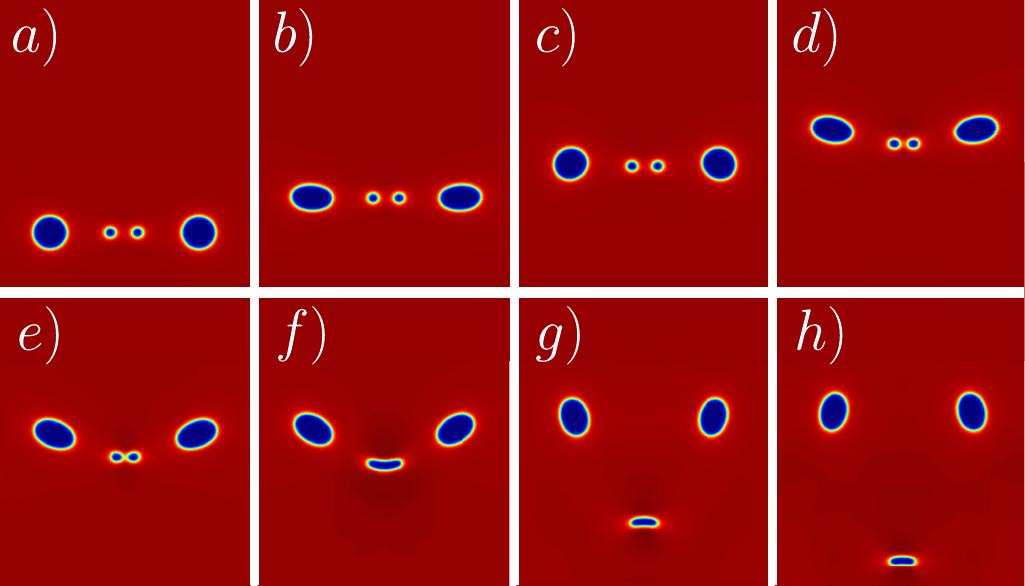}
\end{center}
\caption{
Simulation of the evolution of a vortex array with vortices of $l=\pm 1$ and $\pm 2$, with $L_1=14$ and $L_2$ given
in Eq. (\ref{L2L1}). 
The color code for $|\psi(x,y)|^2$ is the same as that in Figs. \ref{fig2}, \ref{fig5} and the
panels correspond to a) $t=0$, b) $t=1000$, c) $t=2000$, d) $t=3000$, e) $t=3200$, f) $t=3300$, g) $t=3800$, h) $t=4000$, 
The structure moves cohesively with 
the velocity given in Eq. (\ref{UL1val}) up to $t\approx 2500$ and it then gets destabilized. The $l=\pm 2$ vortex-antivortex pair 
continues to move upwards, while the $l=\pm 1$ pair reverses its direction and begins moving backward at a higher velocity.
}
\label{fig9}
\end{figure}

In the second scenario, the $l=\pm 1$ vortices couple to each other and reverse their direction of motion, heading towards negative $x$. Concurrently, the $l=\pm 2$ 
singularities continue moving forward, forming a vortex-antivortex pair similar to those described in Section~\ref{sec:largel},
see Fig.~\ref{fig9}. In summary, the unstable solution is a delicate balance between the ``forces'' that tend to spark both types of instabilities. As shown in Figs. \ref{fig8} and \ref{fig9}, that
balance can be kept for some time if the inter-vortex distances are arranged properly.

%%%%%%%%%%%%%%%%%%%%%% FIGURE 8 %%%%%%%%%%%%%%
%%%%%%%%%%%%%%%%%%%%%%%%%%%%%%%%%%%%%%%%%%

In this contribution, we have not studied solutions of Eqs. (\ref{vipuntual}), (\ref{equalv}) for more than two vortex pairs, $N>2$. In any case, 
we expect any solution based on that structure to be similarly unstable. In general, having more phase singularities would result in more instability modes
for the whole vortex structure.

\section{Conclusions}
\label{sec:conclusion}

In this paper, we have explored, through analytical arguments and numerical computations, a variety of traveling dark excitations in a symmetric, two-dimensional quantum 
droplet medium. These excitations, manifesting as rarefied regions within a liquid-like quantum fluid, can be described as solitonic bubbles moving at constant velocities. 

The study of quantum droplets is a rapidly growing field, encompassing both theoretical and experimental efforts. These novel types of eigenstates provide a fascinating 
addition to the repertoire of phenomena associated with this exotic state of matter. Recent research suggests that the creation of vortices in quantum droplets is 
experimentally feasible and a key question for the next phase of the development in the field \cite{li2024can}. In this context, generating solitonic bubbles could represent 
a natural and promising next step.

Specifically, we have identified a stable branch of solutions that includes configurations such as separated vortex-antivortex pairs, vortex-antivortex pairs confined within 
the same density dip, and rarefaction pulses devoid of phase singularities. Through our numerical results, we have derived relationships among momentum, energy, 
and velocity, and obtained simplified expressions for these quantities in various limits. Additionally, we investigated unstable nonlinear wave solutions, which deform 
during their evolution and eventually disintegrate. These unstable states include vortex-antivortex pairs with topological charges larger than one, as well as finely 
tuned vortex arrays capable of cohesive motion under specific initial conditions.

To conclude, we highlight several open questions inspired by our findings that could guide future theoretical work. While we have focused on stable and unstable 
eigenstates and examined the evolution of instabilities, studying the interactions between these eigenstates could unveil a rich landscape of dynamical
 behaviors \cite{smirnov2015scattering,feijoo2017dynamics}. 
 In particular, it would be interesting to compare the interactions of the vortex dipoles with those found in other types of Bose gases.
 Furthermore, exploring the more realistic scenario of finite droplets would be relevant, particularly in understanding how 
 dark excitations interact with droplet boundaries. Finite droplets might also enable the generation of bubbles through collisions \cite{paredes2014coherent} or 
 support more intricate static vortex configurations \cite{paredes2024self}. 
Finally, extending these studies to alternative models beyond the simple Shannon-type nonlinearity of Eq.~(\ref{eqadim}), such as non-symmetric 
cases \cite{kartashov2025double} or three-dimensional systems \cite{kartashov2018three}, could not only reveal new and unexpected phenomena but 
also bring the theoretical predictions closer to experimental reality.

\section*{Acknowledgements}
We are grateful to three anonymous referees whose insightful comments have significantly improved this manuscript.
This publication is part of the R\&D\&i projects PID2020-118613GB-I00 and PID2023-146884NB-I00,
funded by MCIN/AEI/10.13039/501100011033/. This work was also supported by 
grant ED431B 2024/42 (Xunta de Galicia).

\section*{Data availability statement}

The supporting data for this article, and in particular those used to generate all the figures of the manuscript, are openly available from the Zenodo repository,
reference 15362872. URL: \href{https://doi.org/10.5281/zenodo.15362872}{https://doi.org/10.5281/zenodo.15362872}

\bibliographystyle{apsrev}

\end{document}